\pdfoutput=1

\documentclass{elsarticle}

\usepackage{listings}

\usepackage[colorinlistoftodos,prependcaption,textsize=tiny]{todonotes}

\colorlet{mygray}{black!30}
\colorlet{mygreen}{green!60!blue}
\colorlet{mymauve}{red!60!blue}

\journal{SoftwareX}

\begin{document}

\begin{frontmatter}

\title{GT4Py: High Performance Stencils for Weather and Climate Applications using Python}

\author[cscs]{Enrique G. Paredes}
\author[cscs]{Linus Groner}
\author[htp]{Stefano Ubbiali}
\author[cscs]{Hannes Vogt}
\author[cscs]{Alberto Madonna}
\author[cscs]{Kean Mariotti}
\author[cscs]{Felipe Cruz}
\author[cscs]{Lucas Benedicic}
\author[cscs]{Mauro Bianco}
\author[cscs]{Joost VandeVondele}
\author[cscs,htp]{Thomas C. Schulthess}

\address[cscs]{Swiss National Supercomputing Centre (CSCS), ETH Zurich}
\address[htp]{Institute for Theoretical Physics, ETH Zurich}

\begin{abstract}
All major weather and climate applications are currently developed using languages such as Fortran or C++. This is typical in the domain of high performance computing (HPC), where efficient execution is an important concern. Unfortunately, this approach leads to implementations that intermix optimizations for specific hardware architectures with the high-level numerical methods that are typical for the domain. This leads to code that is verbose, difficult to extend and maintain, and difficult to port to different hardware architectures. Here, we propose a different strategy based on GT4Py (GridTools for Python). GT4Py is a Python framework to write weather and climate applications that includes a high-level embedded domain specific language (DSL) to write stencil computations. The toolchain integrated in GT4Py enables automatic code-generation,to obtain the performance of state-of-the-art C++ and CUDA implementations. The separation of concerns between the mathematical definitions and the actual implementations allows for performance portability of the computations on a wide range of computing architectures, while being embedded in Python allows easy access to the tools of the Python ecosystem to enhance the productivity of the scientists and facilitate integration in complex workflows. Here, the initial release of GT4Py is described, providing an overview of the current state of the framework and performance results showing how GT4Py can outperform pure Python implementations by orders of magnitude.
\end{abstract}

\begin{keyword}
Python \sep stencil \sep weather \sep climate \sep library \sep embedded DSL

\end{keyword}

\end{frontmatter}

\section{Motivation and significance}
\label{sec:motivation}

Simulating weather and climate is a challenging high-performance computing problem with a long history. The simulation requires solving well established differential equations, and also incorporates empirical knowledge and parameterizations to model the subgrid-scale phenomena being considered. These weather models are often part of complex workflows, needed to produce the scientific insight and commercial products derived from them. There are important steps between formulating the problem as a set of mathematical equations and generating simulation results. In particular, algorithms, methods and programs need to be derived, developed, implemented and validated so that the end users can efficiently and reliably employ the massively parallel and advanced computing systems needed to perform the simulations. Here, we present our approach to tackle these steps, using a contemporary approach that aims at providing the domain scientist a suitable and easy-to-use platform for the development of weather models, yet is able to generate highly efficient code.

Up to now, weather and climate models are typically developed in imperative programming languages, such as Fortran. In imperative programming languages programmers can mix mathematical and algorithmic concerns with performance oriented details related to the particular computer architecture that executes the code. There are several drawbacks in this approach, especially as models become more complex and as the code size becomes substantial. To list a few: a) Software maintenance and refactoring become challenging; b) Integration with workflows around the model is non-trivial and fragile; c) Adaptation to new computer architectures becomes challenging, in particular using a single code for multiple platforms d) Difficulty in using or adapting these complex programs for teaching and research; e) Challenges in finding trained people with the skills needed to develop the model, as these skills must range from an understanding of the physics and the numerical methods, to software engineering and computer architecture.

In this paper, we describe GridTools4Py, or GT4Py for short, which aims at alleviating all the issues mentioned above. GT4Py provides a domain specific language (DSL) embedded in Python, and adopts a high-level declarative programming style. As opposed to imperative style programming, declarative programming allows the programmer to focus on scientific, mathematical, and numerical aspects of the computation, separating out implementation aspects. The GT4Py DSL allows for expressing the numerics of the equations in a finite-difference or finite-volume formulation. We believe that the widespread use of Python and the rich ecosystem of tools and environments around it will facilitate the adoption of GT4Py in the weather and climate community, including its use for teaching and research. Modularity and simplicity of the interface are key design objectives. A Python framework should also integrate more easily in typical workflows used by scientists for research and operation. To reach high performance and to adapt to the diversity of today's computer hardware we leverage on just-in-time compilation (JIT) techniques, by which GT4Py can generate and execute highly efficient code based on GridTools libraries \cite{GTPaper} that handle CPU and GPU implementations of the stencil applications arising from the numerical formulations.
The clear separation in frontend and backend is one example of the application of the concept of \emph{separation of concerns}, which declarative programming enables, and finds application in various other frameworks such as, for example, Tensorflow. GT4Py as described in the following corresponds to our first public release of the framework, with a well defined and narrow scope. This version of GT4Py restricts itself to Cartesian grids on regular domains, and single node (CPU and GPU) execution. It serves as a proof-of-concept for an ambitious project to deliver an exascale-ready framework for weather and climate simulations.

\section{Software description}
\label{sec:description}
\subsection{Introduction}

GT4Py supports common numerical methods used in weather and climate applications, namely those derived from finite-difference or finite-volume discretizations on grids\footnote{We use the term ``grid'' here to mean either grid or mesh, since the final computational patterns we describe do not strictly depend on the differences between them.}. Numerical fields are defined on the elements of the grids, and the  algorithms are expressed as applications of functions to a fixed neighborhood of a grid element, whose inputs and outputs are the fields values in that neighborhood. Additionally, in weather and climate applications the vertical direction has usually a special treatment, and the computations can have dependencies on the levels above or below in the atmosphere. In all these cases these computations are  named \emph{stencils}, and are a generalization of the well-known stencil computations on regular grids. 
The iteration space is determined based on the stencil shape and the grid properties.

In GT4Py, stencils are expressed using \emph{GTScript}, a high-level DSL embedded in Python, in which they are encoded inside a Python function and annotated as \lstinline{@gtscript.stencil}. GTScript expressions can be composed to express complex computations, beyond the traditional stencil operators (e.g. laplacians). More details on the definition of GTScript is presented in Section \ref{sec:gtscript}. Syntactically speaking, GTScript is composed by a \emph{strict subset} of the Python language syntax, so it can be parsed by the Python interpreter without requiring further extensions or custom lexer/parser routines. This simplifies substantially the software architecture of GT4Py, and builds on powerful Python functionalities such as \emph{introspection}. The semantics of GTScript statements may differ from the standard Python semantics. This feature allows us to take advantage of the semantics of the domain, to simplify the expression of algorithms, and to generate very efficient code. We note that GT4Py is not a general Python-to-C++ translator, nor a tool to compile and optimize generic Python code, but a domain-specific framework with a precise focus, so that we can attain both domain semantics and efficient implementation.

\subsection{GTScript}
\label{sec:interfaces}\label{sec:gtscript}

\begin{figure}[h!]









\includegraphics[width=\textwidth]{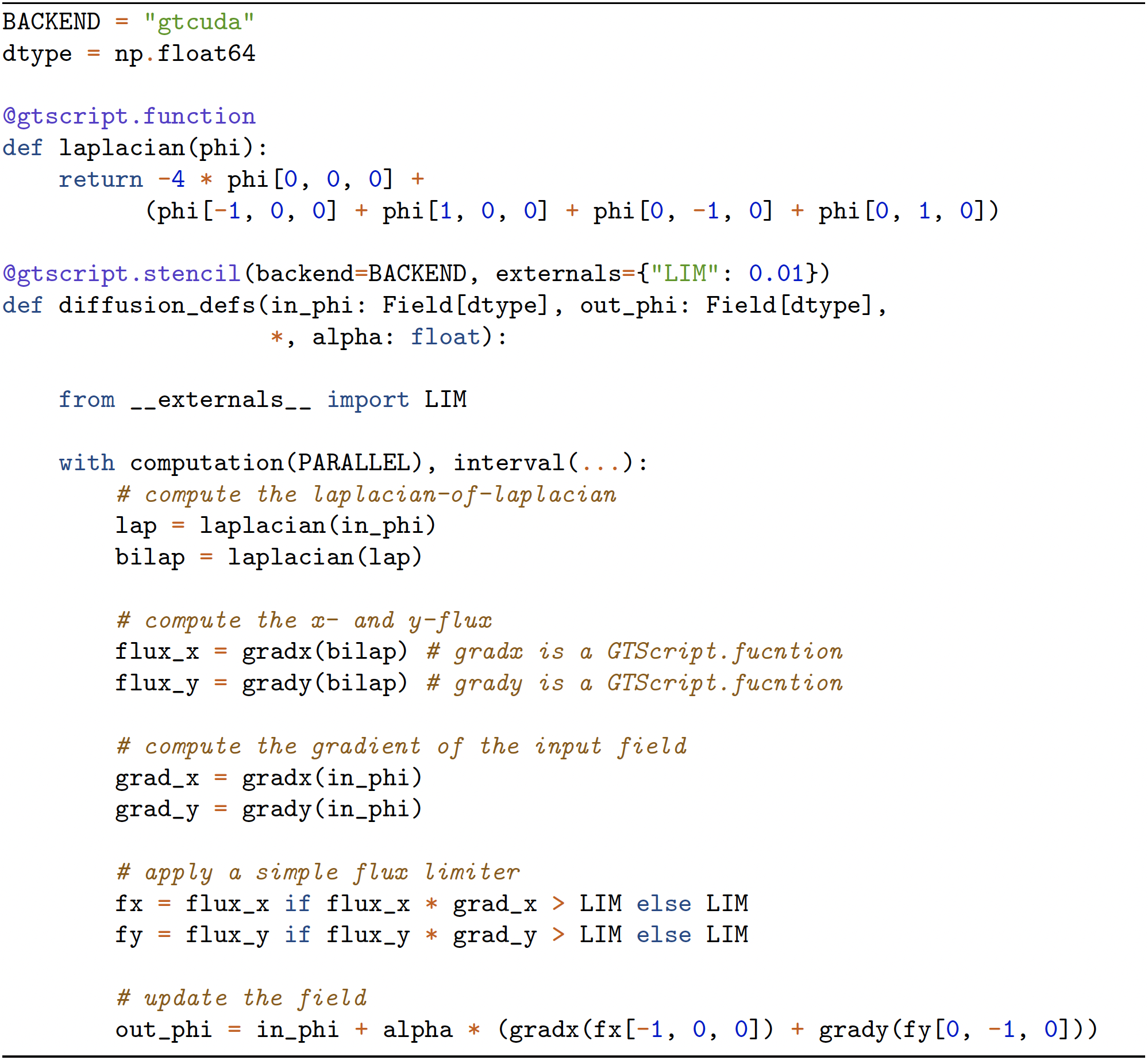}
  \caption{\label{fig:parmodel_gtscript} Examples of a modified tridiagonal-solver on a grid. This code has been designed to highlight some of the most important feature of GT4Py syntax and capabilities: external (inlined) values, non field parameters (\lstinline{a}), temporary fields (\lstinline{tmp}), conditional execution, vertical interval splitting and different execution orders (\lstinline{FORWARD} and \lstinline{BACKWARD}).}
\end{figure}

GTScript is the primary interface provided by GT4Py and it is illustrated in Figure \ref{fig:parmodel_gtscript}. This figure represents a quite complex stencil written using GTScript implementing an horizontal diffusion operator. The code is made complex enough to highlight the main capabilities of the GTScript language. Line 8 shows the main entry point of the computation, where the decorator indicates the backend for which the execution is specified and the definitions of \emph{compile time} constant values, in this case \lstinline{LIM} (these just are two examples of parameters that can be specified in the decorator). Line 9 shows the interface of the function that implements the stencil, which takes two fields and a floating point parameter. The parameters that are not fields are assumed to be read-only scalar parameters, in this case \lstinline{alpha} is an example of those.

After the preamble lines 8-9, the actual computation is specified. The statements in a GTScript function are sequences of \lstinline{with computation(x)}, where \lstinline{x} specifies the iteration order in the vertical direction (\lstinline{PARALLEL}, \lstinline{FORWARD} or \lstinline{BACKWARd}), which ensures that the dependencies in the vertical direction are properly taken into account. These keywords are essential to capture specific features of weather and climate applications. As briefly mentioned above, the execution of the computation is always parallel in the horizontal plane, but can be sequential in the vertical direction. When multiple \lstinline{with computation(x)} statements are specified, in one function body, their execution is equivalent to executing them sequentially in program order, even though the actual execution might be fused or otherwise reorganized for performance reasons.

Next, each \lstinline{with computation} statement contains a sequence of \lstinline{with interval} statements, also assumed to be executed in program order (the with can be skipped if there's only one, like in Figure~\ref{fig:parmodel_gtscript}). This allows the vertical axis to be partitioned so that the computation at certain levels of the atmosphere can be specialized. This is an important feature, for example to specify boundary conditions at the top or bottom layers, or to employ simplified expressions in the upper part of the atmosphere where the influence of the orography is small. The region specification follows the usual Python conventions for ranges, which do not include the end of the range (`None` is used therefore to include the full axis).

Finally, the actual computation can be specified. Currently, GTScript supports only two kind of statements in the body of \lstinline{with interval}: assignments and if-else control-flow operations. These statements have dedicated semantics, since they specify \emph{the application of the expression to all points in the iteration space}, and can be seen as the innermost loop-body of a loop-nest that executes the stencil operator. In particular, this implies an important difference with Python when it comes to field indexing: the indices inside the brackets are interpreted as offsets relative to the point of evaluation of the stencils in the three coordinates. 

The assignment statement, using the \lstinline{=} symbol, is semantically an assignment of the right-hand side (rhs) expression computed over the full domain prior to assignment to the left-hand side (lhs). In general, this would require the creation of a temporary field, which is unacceptable for performance reasons. For this reason, self assignment is forbidden if the computation is \lstinline{PARALLEL} and has dependencies in any direction. This is why line 15 of Figure \ref{fig:parmodel_gtscript} reads form \lstinline{in_lap} and writes into \lstinline{lap}. In both GTScript stencils and functions the offsets where stencils need to access are indicated within square brackets, as visible in lines 6 and 33. The first two offsets indicate horizontal direction, the third indicates an offset on the vertical direction in the atmosphere. In case of \lstinline{FORWARD} and \lstinline{BACKWARD} computations, these offsets are checked at compilation time to detect mistakes in accessing the values.

Fields appearing for the fist time on the lhs of expressions, as in lines 17 to 30, are considered local field variables and are treated as \emph{temporary fields}. This gives GT4Py the ability to exploit the memory systems of the backend architectures, since their values are not observable outside the computations. This is a major feature for reaching high performance, and mark the distinction between the typical imperative programming style from GT4Py. 
Another very important outcome of the GTScript formulation of computations is that the iteration over the domain is implicit. This is essential to simplify the way the user writes code, since computing exact loop-bounds can be tricky, and ultimately also enables performance. 

Finally, Figure \ref{fig:parmodel_gtscript}, line 3 shows another important feature for code reuse and modularity by using \emph{functions}. To be used in GTScript, Python pure functions have to be annotated with \lstinline{@gtscript.function}. 
GT4Py generates a callable Python object implementing the operation defined by the user in the annotated function. The object can be called with the same signature as the annotated function, so it can be called naturally.  The (3D) iteration space is deduced automatically by the field sizes and the stencil shape. However, two optional keyword-only arguments can be provided when calling the generated code: \lstinline{domain} and \lstinline{origin}, to provide more flexibility in specifying the iteration space when needed.

The decorator \lstinline{@gtscript.stencil} can take parameters, the most important of which is the \emph{backend}, that is, the platform where the code will be executed. Backends range from plain Python, a \emph{NumPy} implementation, and a C++ implementation, based on the GridTools libraries that can be executed on CPUs and GPUs. The plain Python backend is useful for debugging purposes, since the generated code can be stepped through, while the NumPy implementation offers somewhat better performance. The best performance is obtained with the C++/GridTools implementation, since it targets specifically performance and provides hardware-specific optimization. The input/output fields are implemented as special NumPy-like multidimensional arrays implementation called \lstinline{storage}. Storage containers can be allocated through the same familiar set of routines used in NumPy for allocation. Additionally, those common functions can take extra parameters. The most important one is the \lstinline{backend} parameter, which customizes the address space, layout, alignment and padding of data storage. 
Since GT4Py storages also implement the so-called Buffer Protocol \cite{BP} sharing these memory buffers with external non-Python libraries is straightforward and overhead-free since it can happen without copies. This provides an important performance benefit for the type of data and computations that are typical in weather and climate simulations.

\subsection{Software Architecture}
\label{sec:architecture}

\begin{figure}[ht]
\centerline{
\includegraphics[width=0.50\textwidth]{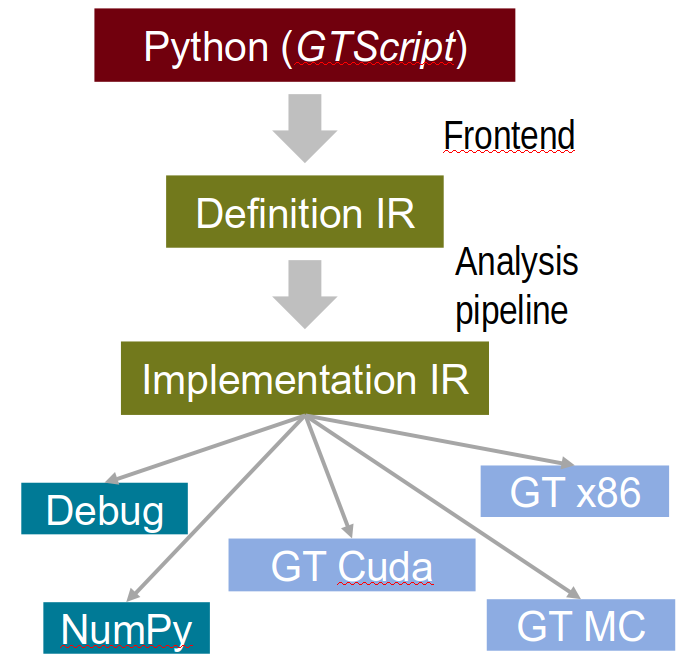}
}
\caption{\label{fig:pipeline} [TODO: redo figure] Shown is the workflow used by GT4Py to generate the code for a given backend. The GTScript functions are transformed into an intermediate representation describing the computation at high level (\emph{definition IR}). Subsequently the analysis pipeline produces the \emph{implementation IR}, which takes into account the parallel execution model and hence can schedule the computations and introduce the necessary synchronizations. From the implementation IR the backend generates the code for the selected executing platform (Python, CUDA, multicore. etc.).}
\end{figure}

GT4Py has a modular and extensible architecture where different execution backends and even DSL frontends can be combined to compose a flexible toolchain. A high-level view of GT4Py is given in Figure~\ref{fig:pipeline}, and shows that GT4Py consists of 
1) frontend components which generate a high-level internal representation of the stencil (definition IR), 
2) an internal analysis pipeline analyzing and transforming this high-level IR into a low-level IR (implementation IR), closer to the parallel model of the implementation, and
3) backend components which generate, compile and load the actual implementations, targeted for a specific architecture or purpose. 

When running a Python model including GT4Py code, GTScript functions will be transparently parsed, analyzed, and transformed into executable code, as the model executes. To avoid the overhead of these operations, both during development and execution, GT4Py provides a caching mechanism to create unique hash identifiers for every stencil implementation. This caching is based on fingerprinting in such a way that code reformatting would not trigger a new compilation. Backends can easily leverage this to avoid the re-compilation of GTScript sources if the code did not change. The key to decouple the different components of the GT4Py toolchain is to use IR objects as an interface between them. These IR objects are trees of simple data classes in a similar spirit to the Python AST, or similar internal representations used in different projects transforming Python to generated code in other languages, like Pythran \cite{guelton2015pythran} or transpyle\cite{transpyle16}.

GT4Py provide, currently, five different backends: \lstinline{debug}, basically provided for debugging purposes, \lstinline{Num} that generates efficient Python code useful for development and prototyping, and finally \lstinline{gtx86}, \lstinline{gtmc} and \lstinline{gtcuda} that generate C++ code based on GridTools APIs to execute the computations natively on multi-cores and GPUs supporting CUDA language.

\section{Impact}
\label{sec:impact}

\subsection{Performance results}

\begin{figure}
   \begin{minipage}{0.49\textwidth}
     \centering
     \includegraphics[width=\linewidth]{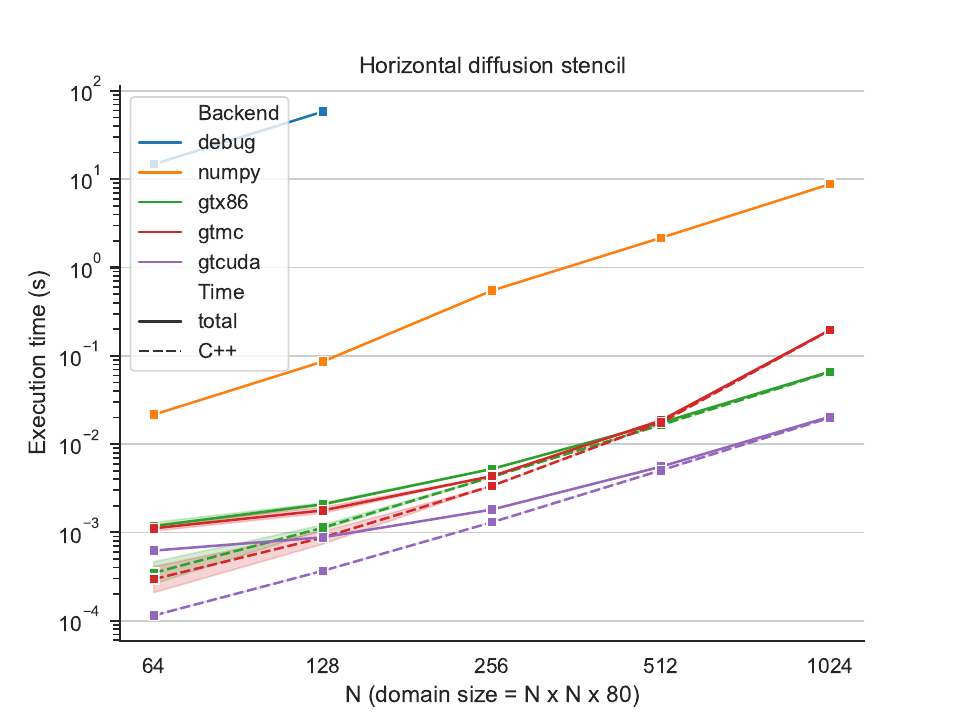}  
   \end{minipage}\hfill
   \begin{minipage}{0.49\textwidth}
     \centering
     \includegraphics[width=\linewidth]{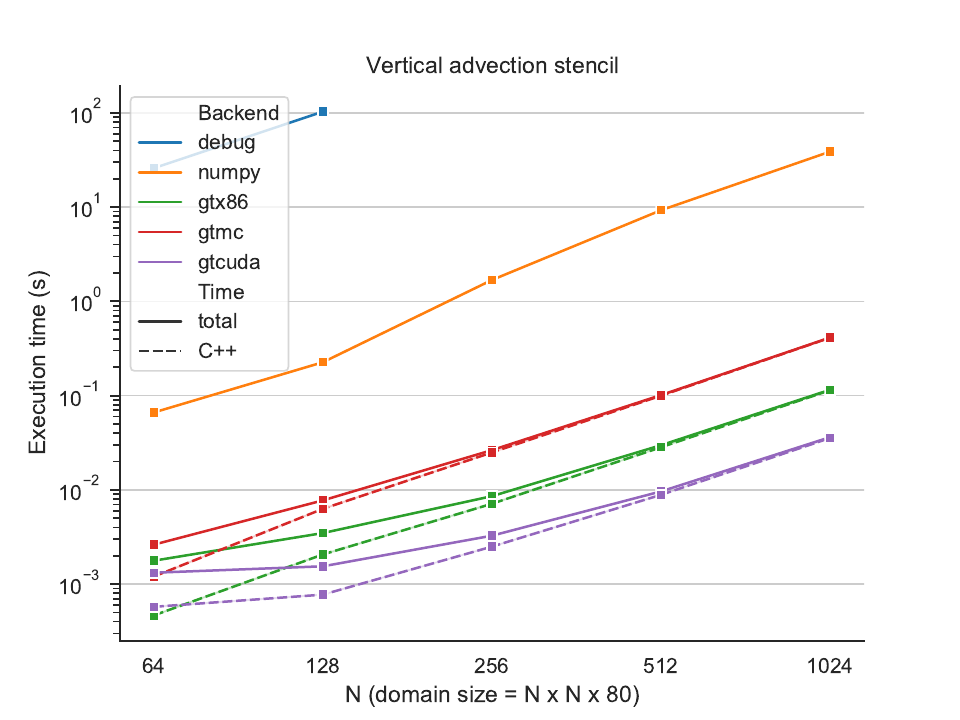}  
   \end{minipage}
\label{fig:hdiff-timings}
\caption{\label{fig:timings}Execution times (log scale) of a parallel and a vertical stencil using different GT4Py backends on a Piz Daint XC50 compute node (Intel Xeon E5-2690 2.60GHz, 12 cores, 64GB RAM CPU and NVIDIA Tesla P100, 16GB RAM GPU). Solid lines show the total call time in the Python interpreter. Dashed lines show the actual execution time of the generated C++ code, skipping safety run-time checks on the provided storage arguments. As expected, the performance optimized GridTools backends are significantly faster than the debug and numpy backends.}
\end{figure}

To evaluate the performance of the different GT4Py backends, we have measured the execution time of two different stencils representing two common patterns in the weather and climate domain: \emph{horizontal diffusion} and \emph{vertical advection}. The horizontal diffusion computation is a classic parallel stencil with several stages and dependencies only in the horizontal plane. Conversely, the vertical advection computation uses different vertical sequential stages to implement an implicit solver for the advection equations.

The results, depicted in Figure \ref{fig:timings}, clearly show that implementations based on GridTools C++ code outperform Python based implementations by a large extent. The slowest C++ implementation running on CPU is at least one order of magnitude faster than the NumPy implementation in both examples. Furthermore, the GPU implementation generated by the \lstinline{gtcuda} backend runs between 5x-10x faster than CPU implementations, depending on the domain size and the particular stencil. These graphs demonstrate that for larger domains, near-native C++ performance is obtained, and thus that advantages of Python as a high-level language can be combined with the performance advantages of C++ as a low-level language. For the most efficient implementations and for small domain sizes, there is a noticeable ($\approx 1$ms) overhead visible as the runtime difference between the overall execution time of the high-level Python function, and the underlying C++ implementations. This constant overhead is caused by various checks performed at run-time on the memory layout and data type of the storage arguments provided by the user. We expect that future versions of GT4Py will provide mechanisms to eliminate this overhead. Finally, it must be emphasized that this performance is obtained using a relatively straightforward analysis pipeline, and we anticipate that progress will be made in generating code that outperforms the currently emitted code.

\subsection{Productivity}
The objective of GT4Py is to bridge the gap between high-productivity programming environments provided by interpreted languages like Python, Matlab, or Julia, and high-performance ones where low-level architectural details can be managed by the programmers for better hardware utilization. In particular, a programmer in Python has access to powerful syntax and a huge set of tools developed by one of the largest communities of developers in the world, with support from non-academic organizations. Similar tools are either not readily available on HPC environments, or with interfaces that are more complicated and more difficult to install/use.

\begin{figure}[ht]
\centerline{
\includegraphics[width=0.9\textwidth]{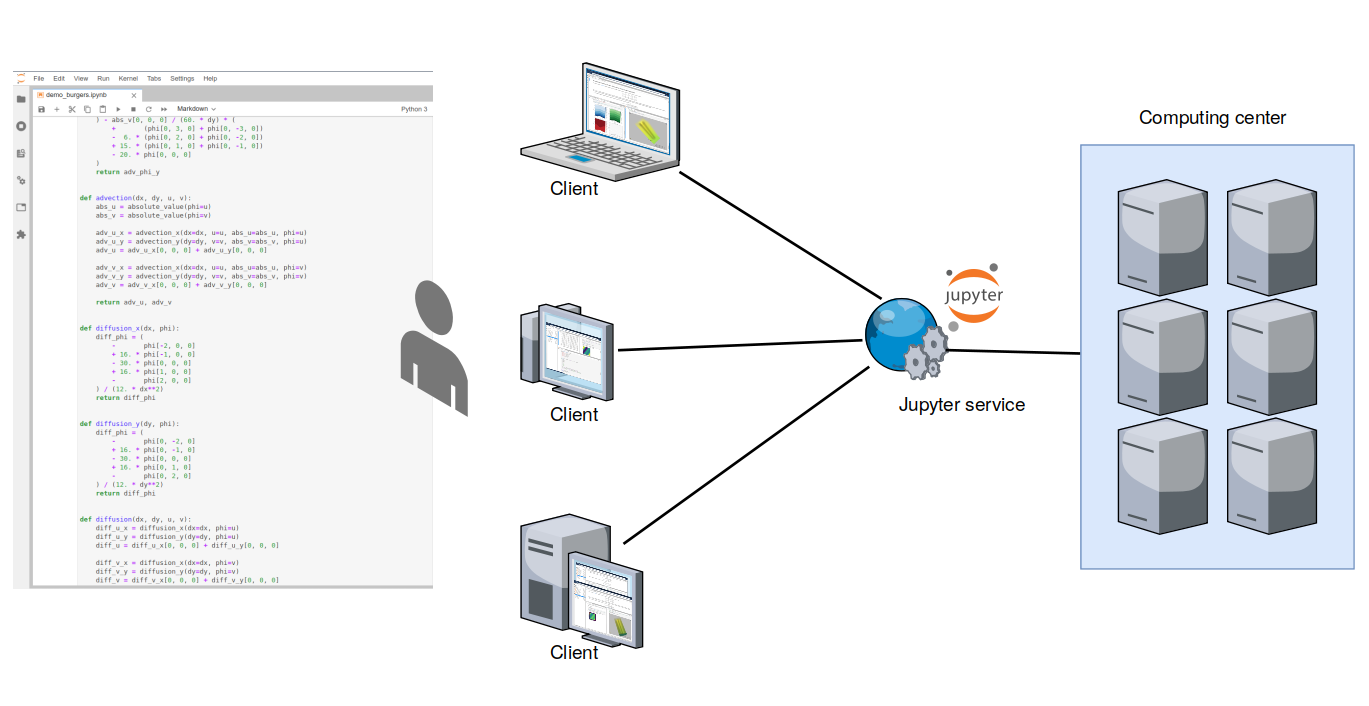}
}
\caption{\label{fig:jupyter} GT4Py can be used from a Jupyter notebook and the code executed on remote computers, in our case the Piz Daint supercomputer at CSCS. This is an example of interactive supercomputing with Jupyter and GT4Py, which represents a shift in the way users can interact with HPC platforms.}
\end{figure}

GT4Py, being embedded in Python, allows the programmers to express all aspects of the simulation at a high level of abstraction, and retains high-performance for the computationally intensive parts. GT4Py does not attempt to be an ontology for the whole application field, but is focused on specific algorithmic motifs. This focus delivers performance, while keeping the toolchain manageable. Using the existing Jupyter \cite{jupyter} infrastructure for interactive supercomputing in Python, it is possible to write GT4Py computations directly in the browser on any local workstation or laptop, and execute a high-performance GPU implementation of the computation directly on the facilities of a supercomputing center (see Fig. \ref{fig:jupyter}). Supporting this kind of simple workflows to access and use computing resources from HPC centers enables rapid progress during the development and prototyping of models. Additionally, this enables the use of these codes for teaching, and facilitates the open publication and exchange of research codes in the scientific community.

\section{Conclusions and Outlook}
\label{sec:conclusions}

We have presented GT4Py, a Python framework for weather and climate applications.  The current version provides an embedded DSL to express stencils on regular (Cartesian) grids, has a functional analysis pipeline, and yields good performance on various computer architectures via the GridTools C++ libraries. This version has been successfully used to develop an isentropic climate model for research purposes, demonstrating that the framework has the features needed to express the numerical patterns of the domain effectively, and demonstrating that it inter-operates with other components of the Python ecosystem. This isentropic model, named Tasmania, will be presented elsewhere. Yet, GT4Py, as presented here, should be considered a prototype of this framework as the functionality is actively extended. 
Ongoing developments will add support for different grids, such as icosahedral, octahedral and cubed-sphere, and will support multi-node (distributed memory) parallelism. This will necessitate extensions of the DSL, adding features and making it more generic, as well as the integration of additional components to deal with multi-node parallelism, such as a halo exchange library\cite{gcl14, GHEX}. Furthermore, also the analysis and backend components of the toolchain will become more powerful, not only to deal with the various grids, but also to perform a wider range of optimizations, and to support additional architectures. Integrating and interfacing to other components, such as data-centric parallel programming environment (DaCe)\cite{dace} and the DAWN compiler\cite{Dawn} are options that are currently currently being explored. These efforts are part of a long term commitment to GT4Py, as GT4Py is part of our ambitious goal of providing and enabling the development of weather and climate models that run effectively on exascale systems and future computer architectures\cite{TCS19}.

\section*{Acknowledgements}
\label{sec:ack}

We thank several individuals and institutions for helping shape the current version of GT4Py. We acknowledge MeteoSwiss, especially Carlos Osuna, for the many discussions about the features needed in weather and climate simulations. We thank ECMWF, in particular Christian K\"uhnlein, and Vulcan Inc., in particular Oliver Fuhrer and his group, for their support and stimulating discussions that helped complete the set of requirements for GT4Py. The feedback and discussions with CSCS colleagues has also been extremely useful in providing the main concepts employed in GT4Py, and gave us confidence in their actual implementation. In particular, we would like to thank Anton Afanasyev, Lukas Mosimann, Felix Thaler, Hannes Vogt and Rico H\"auselmann. This work has been partially funded by the PASC program in Switzerland.


\begin{thebibliography}{00}

\bibitem{TCS19}
T.C.~Schulthess, P.~Bauer, O.~Fuhrer, T.~Hoefler, C.~Schaer, N.~Wedi.
Reflecting on the goal and baseline for exascale computing: a roadmap based on weather and climate simulations. \emph{Computing in Science and Engineering (CiSE)}. Vol 21, Nr. 1, IEEE Computer Society, ISSN: 1521-9615, Jan. 2019.

\bibitem{Gysi15}
T. Gysi, C. Osuna, O. Fuhrer, M. Bianco, and T. C. Schulthess. 
STELLA: A Domain-specific Tool for Structured Grid Methods in Weather and Climate
Models. In Proc. of the Intl. Conf. for High Performance Computing, Networking,
Storage and Analysis (SC ’15). ACM, New York, NY, USA, 2015, Article 41, 12 pages.

\bibitem{Fuhrer14} 
O. Fuhrer and others. 2014. Towards a performance portable, architecture agnostic implementation strategy for weather and climate models.
Supercomputing frontiers and innovations 1, 1 (2014). http://superfri.org/superfri/article/view/17

\bibitem{Vogt18}
O.~Fuhrer, T.~Chadha, T.~Hoefler, G.~Kwasniewski, X.~Lapillonne, D.~Leutwyler, D.~L\"uthi, C.~Osuna, C.~Sch\"ar, T.C.~Schulthess, H.~Vogt. 
Near-global climate simulation at 1km resolution: establishing a performance baseline on 4888 GPUs with COSMO 5.0. Geoscientific Model Development, 11, 1665-1681, 2018.

\bibitem{gcl14}
M.~Bianco. An interface for halo exchange pattern. 2014. www.prace-ri.eu/IMG/pdf/wp86.pdf

\bibitem{guelton2015pythran}
S.~Guelton, P.~Brunet, M.~Amini, A.~Merlini, X.~Corbillon and A.~Raynaud.
Pythran: Enabling static optimization of scientific python programs. Computational Science \& Discovery, vol. 8, 1, 2015.

\bibitem{transpyle16}
M. Bysiek, A. Drozd and S. Matsuoka, Migrating Legacy Fortran to Python While Retaining Fortran-Level Performance Through Transpilation and Type Hints, PyHPC 2016: 6th Workshop on Python for High-Performance and Scientific Computing @ SC16, Salt Lake City, Utah, United States of America, 2016, pp. 9-18

\bibitem{Verbosio18}
F.~Verbosio, J.~Kardos, M.~Bianco, O.~Schenk.
Highly Scalable Stencil-Based Matrix-Free Stochastic Estimator for the Diagonal of the Inverse. Proceedings of the 30th International Symposium on Computer Architecture and High Performance Computing (SBAC-PAD), 2018, 410-419.

\bibitem{GTPaper}
A.~Afanasyev, M.~Bianco, L.~Mosimann, C.~Osuna, F.~Thaler, H.~Vogt, O.~Fuhrer, J.~VandeVondele, T.C.~Schulthess. GridTools: a Framework for Portable weather and climate Applications. Submitted, 2020.

\bibitem{GTrepo} https://github.com/GridTools/gridtools

\bibitem{GHEX} https://github.com/GridTools/GHEX

\bibitem{Dawn} https://github.com/MeteoSwiss-APN/dawn

\bibitem{padal}
D. Unat, A. Dubey, T. Hoefler, J. Shalf, M. Abraham, M. Bianco, B. L. Chamberlain, R. Cledat, H. C. Edwards, H. Finkel, K. Fuerlinger, F. Hannig, E. Jeannot, A. Kamil, J. Keasler, P. H. J. Kelly, V. Leung, H. Ltaief, N. Maruyama, C. J. Newburn, M. Pericas. Trends in Data Locality Abstractions for HPC Systems. IEEE Transactions on Parallel and Distributed Systems, 2017, {\bf 28}(10), 3007--3020, 2017.

\bibitem{jupyter}
T.~Kluyver, B.~Ragan-Kelley, F.~Pérez, B.E.~Granger, M.~Bussonnier, J.~Frederic, K.~Kelley, J.B.~Hamrick, J.~Grout, S.~Corlay, P.~Ivanov, D.~Avila, S.~Abdalla, C.~Willing, and et al. Jupyter Notebooks -- a publishing format for reproducible computational workflows. ELPUB, 2016


\bibitem{cosmo_model}
G.~Doms, M.~Baldauf. A Description of the Nonhydrostatic Regional COSMO-Model - Part I: Dynamics and Numerics. COSMO - Consortium for Small-Scale Modelling, May 2015, http://cosmo-model.org/content/model/documentation/core/cosmoDyncsNumcs.pdf

\bibitem{thaler}
F.~Thaler, S.~Moosbrugger, C.~Osuna, M.~Bianco, H.~Vogt, A.~Afanasyev, L.~Mosimann, O.~Fuhrer, T.~C.~Schulthess, T.~Hoefler.
Porting the COSMO Weather Model to Manycore CPUs.
PASC '19: Proceedings of the Platform for Advanced Scientific Computing Conference, June 2019, 1–-11

\bibitem{dace}
B. Tal, J. de~Fine~Licht, A.N. Ziogas,T. Schneider, T. Hoefler. Stateful Dataflow Multigraphs: A Data-Centric Model for Performance Portability on Heterogeneous Architectures, 2019,Proceedings of the International Conference for High Performance Computing, Networking, Storage and Analysis, SC'19.

\bibitem{BP}
PEP 3118 -- Revising the buffer protocol. https://www.python.org/dev/peps/pep-3118/

\bibitem{GT}
GridTools repository. https://github.com/GridTools/gridtools

\end{thebibliography}
\end{document}